\documentclass[11pt]{article}
\usepackage{graphicx,fullpage,amssymb,url,hyperref} 
\usepackage{orcidlink}

\graphicspath{{Fig/}{./}}
\title{Symplectic Bregman divergences}

\author{Frank Nielsen\textsuperscript{\orcidlink{0000-0001-5728-0726}}\\ \ \\ Sony Computer Science Laboratories Inc.\\ Tokyo, Japan}

\date{(preliminary version)}

\def\calF{{\mathcal{F}}}
\def\natpair#1#2{{(( #1,#2 ))}}
\def\dualprod#1#2{{b(#1,#2)}}
\def\inner#1#2{{\langle #1,#2\rangle}}
\def\cinner#1#2{{\langle\langle #1,#2\rangle\rangle}}
\def\GL{{\mathrm{GL}}}
\def\barF{{\bar{F}}}
\def\barG{{\bar{G}}}
\def\bartheta{{\bar{\theta}}}
\def\Sp{{\mathrm{Sp}}}
\def\bbR{\mathbb{R}}

\def\mattwo#1#2#3#4{{\left[\begin{array}{ll}#1 & #2 \cr  #3 & #4 \end{array}\right]}}
\def\st{{\ :\ }}
\def\dx{\mathrm{d}x}
\def\dy{\mathrm{d}y}
\def\dt{{\mathrm{d}t}}
\def\bbC{\mathbb{C}}
\def\st{{\ :\ }}
\def\dq{{ \mathrm{d}q}}
\def\dpp{{\mathrm{d}p}}  

\def\matrixtwo#1#2#3#4{{\left[\begin{array}{ll}#1 & #2 \cr  #3 & #4 \end{array}\right]}}
\def\calD{\mathcal{D}}
\def\rev{\mathrm{rev}}
\def\irr{\mathrm{irr}}

\newtheorem{Definition}{Definition}
\newtheorem{Theorem}{Theorem}

\newtheorem{Remark}{Remark}
\newtheorem{Property}{Property}

\begin{document}

\maketitle

\begin{abstract}
We present a generalization of Bregman divergences in symplectic vector spaces that we term symplectic Bregman divergences.
Symplectic Bregman divergences are derived from a symplectic generalization of the Fenchel-Young inequality which relies on the notion of symplectic subdifferentials.
The  symplectic Fenchel-Young inequality is obtained using the symplectic Fenchel transform
 which is defined with respect to the symplectic form. 
Since symplectic forms can be generically built from pairings of dual systems, we get a generalization of Bregman divergences in dual systems obtained by equivalent symplectic Bregman divergences. 
In particular, when the symplectic form is derived from an inner product, we show that the corresponding symplectic Bregman divergences amount to ordinary Bregman divergences with respect to composite inner products.
Some potential applications of symplectic divergences in geometric mechanics, information geometry, and learning dynamics in machine learning are touched upon.
\end{abstract}

\noindent Keywords: dual system; duality product; inner product; linear symplectic form; symplectic matrix group; symplectic subdifferential; symplectic Fenchel transform;  Moreau's proximation; geometric mechanics.

\section{Introduction}

\begin{figure}
\centering
\includegraphics[width=0.75\textwidth]{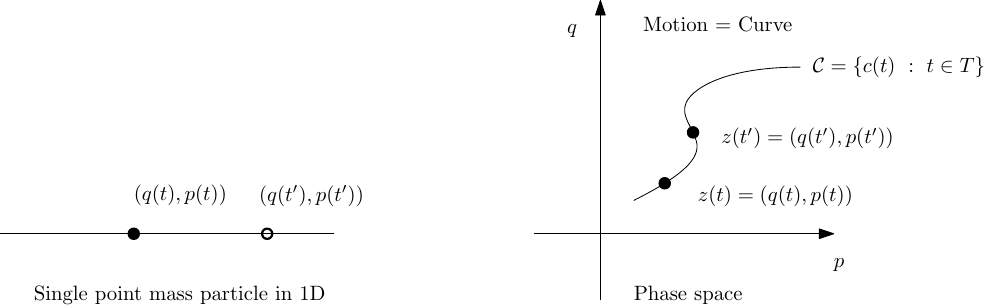}

\caption{The motion of a single point particle $q(t)$ with mass $m$ and momentum $p(t)=m\dot{q}(t)$ on a 1D line can be modeled as a curve $\mathcal{C}=\{c(t)=(q(t),p(t)) \st t\in T\subset\bbR\}$ in the phase space $\bbR^2$.
}\label{fig:phasespace}
\end{figure}

Symplectic geometry~\cite{mcduff1998symplectic,DaSilva-2001,libermann2012symplectic} was historically pioneered by Lagrange around 1808-1810~\cite{L-1808,L-1810,Marle-2009} where the motions and dynamics (evolution curves) of a finite set of $m$ point mass particles in a time interval $T$ are analyzed in the phase space by a 1D curve 
$\mathcal{C}=\{c(t)=(q_1,p_1,\ldots,q_m,p_m)(t) \st t \in T\subset\bbR\}  \subset \bbR^{2n}$, where $q_i(t)\in\bbR^n$'s denote the point locations at time $t$ and $p_i(t)\in\bbR^n$'s encode the momentum, i.e., $p_i(t)=m_i\dot q_i$ with $\dot q_i=\frac{d}{\dt}q_i(t)$.
See Figure~\ref{fig:phasespace}. 
(Notice that Joseph-Louis Lagrange (1736-1813) was 72 years old in 1808, and is famous for his treatise on analytic mechanics~\cite{Lagrange-1811,Lagrange-2013} published first in french in 1788 when he was 52 years old.)

The Hamiltonian coupled equations~\cite{IntroRie-2012} governing the system motion are written in the phase space as
\begin{eqnarray}\label{eq:HamiltonPhysics}
\frac{\dq^i}{\dt} =\frac{\partial{H}}{\partial p_i},\quad
\frac{\dpp_i}{\dt} = -\frac{\partial H}{\partial q^i},
\end{eqnarray}
where $H(q,p,t)$ is the Hamiltonian describing the system. Lagrange originally started a new kind of calculus, ``symplectic calculus.''
Symplectic geometry can be thought as the first discovered non-Euclidean geometric structure since hyperbolic geometry is usually considered to be first studied by Lobachevsky and  Bolyai  around the 1820's-1930's.
We refer to the paper entitled ``The symplectization of science''~\cite{gotay1992symplectization} for an outreach article on symplectic geometry.

The adverb ``symplectic'' stems from Greek: It means ``braided together'' to convey the interactions of point mass particle positions with their momenta.
Its use in mathematics originated in the work of Hermann Weyl (see \S6 on symplectic groups in~\cite{weyl1946classical}).
Another synonym adverb of symplectic is ``complex'' which has been used to describe braided numbers $z$ of $\bbC=\{z=a+ib \st (a,b)\in\bbR^2\}$. Complex has its etymological root in Latin.
In differential geometry, symplectic structures are closely related to (almost) complex structures on vector spaces and smooth manifolds~\cite{DaSilva-2001}.

In physics, symplectic geometry is not only at the core of classical mechanics (i.e., conservative reversible mechanics) and quantum mechanics~\cite{Souriau-1997}, but has also recently been used to model and study dynamics of systems exhibiting dissipative terms~\cite{BuligaSaxce-2017} which are irreversible.
As a pure geometry, symplectic geometry can be studied on its own by mathematicians, and gave birth to the field of symplectic topology~\cite{audin2014vladimir}.
Thus symplectic geometry can be fruitfully applied to various areas beyond its original domain of geometric mechanics.
For example, symplectic geometry has been considered in machine learning for accelerating numerical optimization methods based on symplectic integrators~\cite{Jordan-2018} and in physics-informed neural networks~\cite{chen2021neural,matsubara2021symplectic} (PINNs).

In this paper, we define {\em symplectic Bregman divergences} (Definition~\ref{def:sympBD}) which recover as special cases Bregman divergences~\cite{Bregman-1967} defined with respect to a composite inner-product.
A Bregman divergence induced by a strictly convex and differentiable (potential) function $F$ (called the Bregman generator) between $x_1$ and $x_2$ of $X$ is defined in~\cite{Bregman-1967} (1967) by
\begin{equation}\label{eq:BD}
B_F(x_1:x_2) = F(x_1)-F(x_2)-\inner{x_1-x_2}{\nabla F(x_2)},
\end{equation}
where $\inner{\cdot}{\cdot}$ is an inner product on $X$.
Let $\Gamma_0(X)$ denote the set of functions which are lower semicontinuous  convex with non-empty effective domains.
The convex conjugate $F^*(x^*)$ obtained by the Legendre-Fenchel transform $F^*(x^*)=\sup_{x\in X} \inner{x^*}{x}-F(x)$ yields a dual Bregman divergence $B_{F^*}$ when the function $F\in\Gamma_0(X)$ is of Legendre type~\cite{Rockafellar-1967,BanachLegendreFunc-2001}:
$$
B_{F*}(x_1*:x_2^*)=F^*(x_1^*)-F^*(x_2^*)-\inner{x_1^*-x_2^*}{\nabla F^*(x_2^*)},
$$
such that $B_F(x_1:x_2)=B_{F^*}(x_2^*:x_1^*)$ with $F^*(x^*)=\inner{x^*}{(\nabla F)^{-1}(x^*)}-F((\nabla F)^{-1}(x^*))$.

This paper introduces and extends the work of Buliga and Saxc\'e~\cite{BuligaSaxce-2017} which is motivated by geometric irreversible mechanics.
To contrast with~\cite{BuligaSaxce-2017}, this expository  paper is targeted to an audience familiar with Bregman divergences~\cite{Bregman-1967} 
in machine learning and information geometry~\cite{IG-2016} but do not assume any prior knowledge in geometric mechanics.

The paper is organized as follows: In \S\ref{sec:form}, we define symplectic vector spaces and explain the representation of symplectic forms using dual pairings.
We then define the symplectic Fenchel transform and symplectic Fenchel-Young inequality in \S\ref{sec:FYineq}.
The definitions of symplectic Fenchel-Young divergences (Definition~\ref{def:sympY}) and symplectic Bregman divergences (Definition~\ref{def:sympBD}) are reported in \S\ref{sec:sympdiv}.
In particular, we show how to recover Bregman divergences with respect to composite inner-products as special cases in \S\ref{sec:compBD} (Property~\ref{prop:bdci}).
In general, symplectic Bregman divergences allow to define Bregman divergences in dual systems equipped with pairing products.
Finally, we recall the role of Bregman divergences in dually flat manifolds of information geometry in \S\ref{sec:concl}, and motivate the introduction of symplectic Bregman divergences in geometric mechanics (e.g., symplectic BEN principle of~\cite{BuligaSaxce-2017}) and learning dynamics in machine learning. 

\section{Dual systems, symplectic forms, and symplectomorphisms}\label{sec:form}

\subsection{Symplectic forms derived from dual systems}

We begin with two definitions:

\begin{Definition}[Dual system.]\label{def:ds}
Let $X$ and $Y$ be topological locally convex $m$-dimensional vector spaces~\cite{Horvath-2013} (LCTVSs) equipped with a pairing product $\dualprod{\cdot}{\cdot}$, i.e., a bilinear map:
$$
\dualprod{\cdot}{\cdot}: X\times Y\rightarrow\bbR,
$$
such that all continuous linear functionals on $X$ and $Y$ are expressed as $x^{\#}(\cdot)=\dualprod{x}{\cdot}$ and $y{^\#}(\cdot)=\dualprod{\cdot}{y}$, respectively. 
The triplet $(X,Y,\dualprod{\cdot}{\cdot})$ forms a dual system.
\end{Definition}

(Notice that when the type of $X$ is different from the type of $Y$ then the bilinear map cannot be symmetric.)

\begin{Definition}[Symplectic vector space]
A symplectic vector space $(V,\omega)$ is a vector space equipped with a map~\cite{McInerney-2013} $\omega: Z=V\times V\rightarrow\bbR$ which is
\begin{enumerate}
\item bilinear: $\forall \alpha,\beta,\alpha',\beta'\in\bbR, \forall z_1,z_2\in Z$, we have
$$
\omega(\alpha z_1+\alpha' z_1',z_2,\beta z_2+\beta' z_2')=
\alpha\beta\,\omega(z_1,z_2)
+\alpha\beta'\,\omega(z_1,z_2')+
\alpha'\beta\,\omega(z_1',z_2)+
\alpha'\beta'\,\omega(z_1,z_2'), 
$$

\item skew-symmetric (or alternating): $\omega(z_2,z_1)=-\omega(z_1,z_2)$, and

\item non-degenerate: if for a $z_0$, we have $\omega(z,z_0)=0$ for all  $z\in Z$ then we have $z_0=0$.
\end{enumerate}
 \end{Definition}

Notice that skew-symmetry implies that $\omega(z,z)=0$ for all $z\in Z$ since $\omega(z,z)=-\omega(z,z)$ and hence $2\omega(z,z)=0$. 
The map $\omega$ is called a linear symplectic form~\cite{McInerney-2013,Bourguignon-2022}.

We define the symplectic form $\omega$ induced by the pairing product of a dual system as follows:
\begin{equation}
\omega(z_1,z_2) = \dualprod{x_1}{y_2}-\dualprod{x_2}{y_1},
\end{equation}
where $z_1=(x_1,l_1)$ and $z_2=(x_2,l_2)$ belong to $Z=V\oplus V^*$.

Let us report several examples of linear symplectic forms:

\begin{itemize}

\item Let $X=V$ be a finite $n$-dimensional vector space with the dual space of linear functionals  $Y=V^*$ (space of covectors $l$).
 The natural pairing $\natpair{x}{l}=l(x)=\sum_i x^il_i$ of a vector $x\in V$ with a covector $l\in V^*$ is an example of dual product.
(We use the superscript index for indicating components of contravariant vectors and subscript index for specifying components of covariant vectors~\cite{IntroRie-2012}.)
We define the symplectic form $\omega$ induced by the natural pairing of vectors with covectors as follows:
\begin{equation}
\omega(z_1,z_2) = \natpair{x_1}{y_2}-\natpair{x_2}{y_1} = l_2(x_1)-l_1(x_2),
\end{equation}
where $z_1=(x_1,l_1)$ and $z_2=(x_2,l_2)$ belong to $Z=V\oplus V^*$.
 
\item Consider $(X,\inner{\cdot}{\cdot})$ an inner product space of dimension $n$.
The product space $Z=X\oplus X$ of even dimension $n$ can be equipped with the following map 
$\omega:Z\times Z\rightarrow\bbR$ induced by the inner product:
\begin{equation}\label{eq:omega}
\omega(z_1,z_2)=\inner{x_1}{y_2}-\inner{x_2}{y_1},
\end{equation}
where $z_1=(x_1,y_1)\in Z$ and $z_2=(x_2,y_2)\in Z$.

\begin{figure}
\centering
\includegraphics[width=0.65\textwidth]{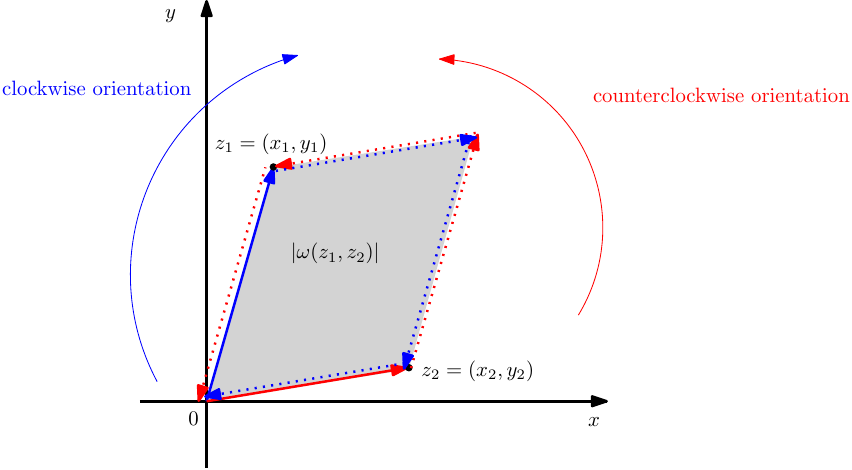}

\caption{Interpreting a 2D symplectic form $\omega(z_1,z_2)$ as the signed area of a parallelogram  with first oriented edge $z_1$ (grey).
A pair of vectors defines two possible orientations of the parallelogram: The orientation compatible with $z_1$ and the reverse orientation compatible with $z_2$.  
$\omega$ is called the standard area form. }\label{fig:warea}
\end{figure}

For example, let $X=\bbR$ and $\inner{x_1}{x_2}=x_1x_2$.
Then $\omega(z_1,z_2)={x_1}{y_2}-{x_2}{y_1}$.
This symplectic form can be interpreted as the determinant of the matrix $M=\mattwo{x_1}{x_2}{y_2}{y_1}$ which corresponds geometrically to the signed orientation of the parallelogram defined by the vectors $z_1=(x_1,y_1)$ and $z_2=(x_2,y_2)$. See Figure~\ref{fig:warea}. 
(This example indicates the link with integration of 2D manifolds equipped with fields of symplectic forms smoothly varying called differential $2$-forms~\cite{IntroRie-2012}.) 

In a finite-dimensional vector space, we can express the inner product as $\inner{x}{y}=x^\top Q y$ for a symmetric positive-definite matrix $Q\in \bbR^{n\times n}$.
Let $Q=L^\top L$ be the Cholesky decomposition of $Q$. Then we have 
$$
\inner{x}{y}= (L^\top x)^\top I (L^\top y)=\inner{L^\top x}{L^\top y}_0,
$$
 where $I$ is the $n\times n$ identity matrix and $\inner{x}{y}_0=x^\top y$ is the Euclidean inner product.
Thus the form $\omega_0$ induced by $\inner{\cdot}{\cdot}_0$ can be expressed using linear algebra as
$$
\omega_0(z_1,z_2)= z_1^\top \matrixtwo{0}{I}{-I}{0} z_2=z_1^\top \Omega_0 z_2,
$$
where $\Omega_0\in\bbR^{2n\times 2n}$ is a skew-symmetric matrix: $\Omega_0^\top=-\Omega_0$.
More generally, we may consider skew-symmetric matrices of the form $\Omega=\matrixtwo{0}{L^\top}{-L^\top}{0}$ to define the symplectic form $\omega_Q$ induced by the inner product $\inner{x}{y}_Q=x^\top Q y$.

\item Let $(X,\|\cdot\|)$ be a reflexive real Banach space with topological dual $(Y=X^*,\|\cdot\|_*)$ and canonical bilinear form $\dualprod{\cdot}{\cdot}$.
Then $(X,X^*,\dualprod{\cdot}{\cdot})$ is a dual system.

\end{itemize}

\subsection{Linear symplectomorphisms and the groups of symplectic matrices}

A symplectic form $\omega$ can be expressed as a $2n\times 2n$ matrix $\Omega=[\omega_{ij}]$ such that
 $\omega_{ij}=\omega(b_i,b_j)$ where $b_1=e_1,\ldots, b_n=e_n,b_{n+1}=f_1,\ldots,b_{2n}=f_n$ are the basis vectors, 
and $\omega(z_1,z_2)=z_1^\top \Omega z_2$.

The Darboux basis~\cite{DaSilva-2001} of the canonical form $\omega_0$ of $\bbR^{2n}$ is such that 
 $\omega(e_i,f_j)=\delta_{ij}$ and $\omega(e_i,e_j)=\omega(l_i,l_j)=0$ where $\delta$ denotes the Dirac function. 
$\Omega_0\in\Sp(2n)$ is the symplectic matrix $\matrixtwo{0}{I}{-I}{0}$ corresponding to the canonical form $\omega_0$ of $\bbR^{2n}$.

A transformation $t:V\rightarrow V$ is called a linear symplectomorphism when $\omega(t(z_1),t(z_2)=\omega(z_1,z_2)$ (i.e., $t^*\omega=\omega$), i.e.,
 when $T^\top\Omega T=\Omega$ where $T$ be the matrix representation of $t$. 
In particular $t$ is a linear symplectomorphism with respect to $\omega_0$ when $T^\top \Omega_0 T =\Omega_0$.
Any symplectic vector space of $(V,\omega)$ dimension $2n$ is symplectomorphic to the canonical symplectic space $(\bbR^{2n},\omega_0)$.

Linear symplectomorphisms can be represented by symplectic matrices of the symplectic group~\cite{weyl1946classical,Siegel-1964} $\Sp(2n)$:
\begin{eqnarray*}
\Sp(2n) &=& \left\{T \st T^\top \Omega_0 T =\Omega_0\right\}\subset \mathrm{GL}(2n),\\
&=&  \left\{ T=\matrixtwo{A}{B}{C}{D} \st -C^\top A+A^\top C=0, -C^\top B+A^\top D=I, -D^\top B+B^\top D=0 \right\},
\end{eqnarray*}
 
Transpose and inverse of symplectic matrices are symplectic matrices.
The inverse of a symplectic matrix $T$ is given by
\begin{eqnarray*}
T^{-1} &=& -\Omega_0 T^\top \Omega_0,\\
&=& \matrixtwo{D^\top}{-B^\top}{-C^\top}{A^\top}
\end{eqnarray*}

Symplectic matrices of $\Sp(2n)$ have unit determinant ($\Sp(2n)\subset\mathrm{SL}(2n)\subset\mathrm{GL}(2n)$), and in the particular case of $n=1$, $\Sp(2)$ corresponds precisely to the set of matrices with unit determinant. 
Thus rotation matrices of $\mathrm{SO}(2)$ which have unit determinant for a subgroup of $\Sp(2)$.

Sesquilinear symplectic forms can also be defined on  complex linear spaces~\cite{everitt1999complex}.

\section{Symplectic Fenchel transform, symplectic subdifferentials, and symplectic Fenchel-Young (in)equality}\label{sec:FYineq}

Let $F:Z=X\times Y\rightarrow \bbR\cup\{+\infty\}$ be a convex lower semi-continuous (lsc) function called a potential function.
The symplectic Fenchel conjugate $F^{*\omega}(z')$  is  defined by
$$
F^{*\omega}(z') = \sup_{z\in Z} \left\{ \omega(z',z)-F(z) \right\}.
$$

Notice that since $\omega$ is skew-symmetric, the order of the arguments in $\omega$ is important: The 
symplectic Fenchel transform optimizes with respect to the second argument.

\begin{Remark}
Moreau generalized the Fenchel conjugate using a cost function~\cite{moreau1970inf}. 
In particular, the duality induced by logarithmic cost function was studied in~\cite{wong2018logarithmic}, and lead to a generalization of Bregman divergences called the logarithmic divergences which are canonical divergences of constant section curvature manifolds in information geometry.
\end{Remark}

The symplectic subdifferential of $F$ at $z$ is defined by
$$
\partial^\omega F(z)=\left\{
z_1\in Z \st \forall z_2\in Z, \quad
F(z+z_2) \geq F(z)+\omega(z_1,z_2)
\right\}.
$$

The differential operator $\partial^\omega$ is a set-valued operator: $\partial^\omega: \calF \rightrightarrows Z$, where $\calF$ is the set of potential functions.
An element of the symplectic subdifferential of $F$ at $z$ is called a symplectic subgradient.

\begin{Remark}
In geometric mechanics~\cite{DaSilva-2001}, the symplectic gradient on a symplectic manifold $(M,\omega)$ is the Hamiltonian vector field, i.e., the vector field $X_H$such that the Halmitonian mechanics equation writes concisely as $\omega(X_H,\cdot)=\mathrm{d}H$.
\end{Remark}

\begin{Theorem}[Symplectic Fenchel-Young inequality, Theorem 2.3 of~\cite{BuligaSaxce-2017}]\label{thm:sympFY}
Let $F(z)$ be a convex (i.e., $F(z)=F(x,y)$ is joint convex, i.e., convex with respect to $z=(x,y)$) and lower semicontinuous function. 
Then the following inequality holds:
$$
\forall z,z'\in Z,\quad F(z)+F^{*\omega}(z')\geq \omega(z',z),
$$
with equality if and only if $z'\in\partial^\omega(z)$.
\end{Theorem}

Let us again notice that the argument order in $\omega$ is important.

Assume that the potential functions are smooth and that symplectic differentials consist only of single-element sets (singletons).
By abuse of language, we shall call in this paper the symplectic gradient of $F$ the single element of the symplect subdifferential $\partial^\omega$,
 and denote it by $\nabla^\omega F$: $\partial^\omega F(z)=\{\nabla^\omega F(z)\}$.
(Our terminology and notation is thus  not to be confused with the Hamiltonian vector field $X_H$ of geometric mechanics.)
 
\section{Symplectic Fenchel-Young divergences and symplectic Bregman divergences}\label{sec:sympdiv}

Divergences are smooth dissimilarity functions (see \S4.2 of~\cite{leok2017connecting}).
From the symplectic Fenchel-Young inequality of Theorem~\ref{thm:sympFY}, we can define the symplectic Fenchel-Young divergence as follows:

\begin{Definition}[Symplectic Fenchel-Young divergence]\label{def:sympY}
Let $F:Z=X\times Y\rightarrow\bbR$ be a smooth convex function.
Then the symplectic Fenchel-Young divergence is the following non-negative measure of dissimilarity between $z$ and $z'$:
\begin{equation} 
Y_{F}(z,z') = F(z)+F^{*\omega}(z') - \omega(z',z)\geq 0. \label{eq:sympY}
\end{equation}
\end{Definition}

We have $Y_{F}(z,z')=0$ if and only if $z'\in\partial^\omega F(z)$, i.e., $z'=\nabla^\omega F(z)$ when $F$ is smooth.

Let us now define the symplectic Bregman divergence $B_F^\omega(z_1:z_2)$ 
as $Y_{F}(z_1,z'_2)$ where $z'_2=\nabla^\omega F(z_2)$.
Using the following identity derived from the symplectic Fenchel-Young equality:
$$
F^{*\omega}(\nabla^\omega F(z))=\omega(\nabla^\omega F(z),z)-F(z),
$$
 and the bilinearity of the symplectic form, we get:

\begin{eqnarray}
B_F^\omega(z_1:z_2) &=& Y_{F}(z_1,z'_2),\nonumber\\
&=& F(z_1) + F^{*\omega}(z'_2) - \omega(z'_2,z_1),\nonumber\\
&=& F(z_1) + \omega(\nabla^\omega F(z_2),z_2)-F(z_2) - \omega(\nabla^\omega F(z_2),z_1),\nonumber\\
&=& F(z_1)-F(z_2)-\omega(\nabla^\omega F(z_2),z_1-z_2).\label{eq:sympBD1} 
\end{eqnarray}

Since $\omega$ is skew-symmetric, we can also rewrite Eq.~\ref{eq:sympBD1} equivalently  as
\begin{equation}
B_F^\omega(z_1:z_2) = F(z_1)-F(z_2)+\omega(z_1-z_2,\nabla^\omega F(z_2)).\label{eq:sympBD2} 
\end{equation}

\begin{Definition}[Symplectic Bregman divergence]\label{def:sympBD}
Let $(Z,\omega)$ be a symplectic vector space. 
Then the symplectic Bregman divergence between $z_1$ and $z_2$ of $Z$ induced by a smooth convex potential $F(z)$ is
$$
B_F^\omega(z_1:z_2) = F(z_1)-F(z_2) - \omega(\nabla^\omega F(z_2),z_1-z_2),
$$
where the symplectic subdifferential gradient is the singleton $\partial^\omega F(z)=\{\nabla^\omega F(z)\}$.
\end{Definition}

Notice that ordinary Bregman divergences (BDs) have been generalized to non-smooth strictly convex potential functions using a subdifferential map in~\cite{Kiwiel-1997,BD-Gordon-1999,Iyer-2012} to choose among potential several subgradients at a given location.
Similarly, we can extend symplectic Bregman divergences to non-smooth strictly convex potential functions using a symplectic subdifferential map.

\section{Particular cases recover composite Bregman divergences}\label{sec:compBD}

When $Y=X$ and $(X,\inner{\cdot}{\cdot})$ is an inner-product space, we may consider the composite inner-product on $Z=X\times X$: 
$$
\cinner{z_1}{z_2}=\inner{x_1}{y_1}+\inner{x_2}{y_2},
$$
with $z_1=(x_1,y_1)$ and $z_1=(x_2,y_2)$.

Let $I:Z\rightarrow Z$ be the linear function $I(z)=z$ and denote by $J:Z\rightarrow Z$ the linear function defined by
$$
J(z)=J(x,y)=(-y,x).
$$
Notice that this definition of $J$ makes sense because $X=Y$ and thus $(-y,x)\in Z$. 
We check that we have $J^2(x,y)=J(-y,x)=(-x,-y)=-(x,y)$, i.e., $J^2=-I$.
Furthermore, we have $g(z_1,z_2)=\omega(z_1,Jz_2)=\inner{x_1}{x_2}+\inner{y_1}{y_2}$ that is a positive definite inner product.
That is, the automorphism $J$ is a complex structure $\omega$-compatible ($J$ is a symplectomorphism).

We can express the symplectic form $\omega(z_1,z_2)=\inner{x_1}{y_2}-\inner{x_2}{y_1}$ induced by the inner product using the composite inner product as follows:
\begin{eqnarray*}
\omega(z_1,z_2) &=& \inner{x_1}{y_2}-\inner{x_2}{y_1} = \cinner{J(z_1)}{z_2},\\
\omega(-Jz_1,z_2) &=& \cinner{z_1}{z_2}.
\end{eqnarray*}

Similarly, the symplectic subdifferential of $F$ can be expressed using the ordinary subdifferential (and vice-versa) as follows:

\begin{eqnarray*}
z'\in\partial^\omega F(z) &\Leftrightarrow & J(z')\in\partial F(z),\\
z'\in\partial F(x)  &\Leftrightarrow & -J(z')\in \partial^\omega F(z).
\end{eqnarray*}

When subdifferentials are singletons, we thus have
\begin{eqnarray*}
J(\nabla^\omega F(z)) &=& \nabla F(z),\\
\nabla^\omega F(z) &=& -J(\nabla F(z)).
\end{eqnarray*}

Last, the symplectic Fenchel conjugate of $F$ is related by the ordinary Fenchel conjugate $F^*$ of $F$ as follows:
$$
F^{*\omega}(z)=F^*(J(z)).
$$

Thus in that case the symplectic Bregman divergence amounts to an ordinary Bregman divergence:
\begin{eqnarray*}
B^\omega_F(z_1:z_2) &=& F(z_1)-F(z_2)-\omega(\nabla^\omega F(z_2),z_1-z_2),\\
&=& F(z_1)-F(z_2)+\omega(-\nabla^\omega F(z_2),z_1-z_2),\\
&=& F(z_1)-F(z_2)+\omega(J(\nabla F(z_2)),z_1-z_2),\\
&=& F(z_1)-F(z_2) -\cinner{z_1-z_2}{\nabla F(z_2)},\\
&=& B_F(z_1:z_2).
\end{eqnarray*}

\begin{Property}\label{prop:bdci}
When the symplectic form $\omega$ is induced by an inner-product $\inner{\cdot}{\cdot}$ of $X$, the symplectic Bregman divergence 
$B^\omega_F(z_1:z_2)$ between $z_1=(x_1,y_1)$ and $z_2=(x_2,y_2)$ of $Z=X\times X$ amounts to an ordinary Bregman divergence with respect to the composite inner-product 
$\cinner{z_1}{z_2}=\inner{x_1}{y_1}+\inner{x_2}{y_2}$:
$$
B^\omega_F(z_1:z_2) =B_F(z_1:z_2) = F(z_1)-F(z_2) -\cinner{z_1-z_2}{\nabla F(z_2)}.
$$
\end{Property}

Furthermore, if the potential function $F(z)$ is separable, i.e., $F(z)=F_1(x)+F_2(y)$ for Bregman generators $F_1$ and $F_2$,
then we have $B^\omega_F(z_1:z_2)=B_{F_1}(x_1:x_2)+B_{F_2}(y_1,y_2)$ where the Bregman divergences $B_{F_1}$ and $B_{F_2}$ are defined with respect to the inner product of $X$. 

Notice that the symplectic Fenchel-Young inequality can be rewritten using the ordinary Fenchel-Young inequality and the linear function $J$ as:
\begin{eqnarray*}
F(z)+F^{*\omega}(z') &\geq& \omega(z',z),\\
F(z)+F^*(J(z')) &\geq& \cinner{J(z')}{z},\\
F(x,y)+F^*(-y',x') &\geq & \cinner{(-y',x')}{(x,y)},\\
F(x,y)+F^*(-y',x') &\geq & \inner{x'}{y}-\inner{x}{y'}.
\end{eqnarray*}

\section{Summary, discussion, and perspectives}\label{sec:concl}

Since it inception in operations research, Bregman divergences~\cite{Bregman-1967} have proven instrumental in many scientific fields including information theory, statistics, and machine learning, just to cite a few. 
Let $(X,\inner{\cdot}{\cdot})$ be a Hilbert space, and $F:X\rightarrow\bbR$ a strictly convex and smooth realvalue function.
Then the Bregman divergence induced by $F$ is defined in~\cite{Bregman-1967} (1967) by
$$
B_F(x_1:x_2)=F(x_1)-F(x_2)-\inner{x_1-x_2}{\nabla F(x_2)}.
$$

In information geometry~\cite{IG-2016,EIG-2020,IG-2022}, a smooth dissimilarity $\calD(p,q)$ between two points $p$ and $q$ on a $n$-dimensional smooth manifold $M$  induces a statistical structure on the manifold~\cite{Eguchi-1985}, i.e., a triplet $(g,\nabla,\nabla^*)$ where the Riemannian metric tensor $g$ and the torsion-free affine connections $\nabla$ and $\nabla^*$ are induced by the divergence $\calD$. The duality in information geometry is expressed by the fact that the mid-connection $\frac{\nabla+\nabla^*}{2}$ corresponds to the Levi-Civita connection induced by $g$. 
To build the divergence-based information geometry~\cite{amari2010information}, the divergence $\calD(p,q)$ is interpreted as a scalar function on the product manifold $M\times M$ of dimension $2n$.
Thus the divergence $\calD$ is called a contrast function~\cite{Eguchi-1985} or yoke~\cite{Barndorff-1997}.
Conversely, a statistical structure $(g,\nabla,\nabla^*)$ on a $n$-dimensional manifold $M$ induces a contrast function~\cite{Matumoto-1993}.
When the statistical manifold $(M,g,\nabla,\nabla^*)$ is dually flat with $\theta(\cdot)$ the global $\nabla$-affine coordinate system and $\eta(\cdot)$ the global $\nabla^*$-affine coordinate system~\cite{Shima-2007}, there exists two dual global potential functions $\phi$ and $\phi^*$ on the manifold $M$ such that $\phi(p)=F(\theta(p))$ and $\phi^*(\eta(p))=F^*(\eta(p))$ where $F^*(\eta)$ is the Legendre-Fenchel convex conjugate of $F(\theta)$. 
The canonical dually flat divergence on $M$ is then defined by
$$
\calD(p,q)=F(\theta(p))+F^*(\eta(p))-\sum_{i=1}^n \theta_i(p)\eta_i(q),
$$ 
and amounts to a Fenchel-Young divergence or equivalently a Bregman divergence: 
$$
\calD(p,q)=Y_F(\theta(p):\theta(q))=B_F(\theta(p):\theta(q)),
$$
where the Fenchel-Young divergence is defined by
$$
Y_F(\theta:\eta')=F(\theta)+F^*(\eta')-\sum_{i=1}^n \theta^i\eta'_i.
$$
The Riemannian metric $g$ of a dually flat space can be expressed as $g=\nabla\mathrm{d}\phi=\nabla^*\mathrm{d}\phi^*$ or in the $\theta$-coordinates by
$g_{ij}(\theta)=\frac{\partial^2}{\partial\theta_i\partial\theta_j} F(\theta)$ and in the $\eta$-coordinates by 
$g_{ij}(\eta)=\frac{\partial^2}{\partial\eta_i\partial\eta_j} F^*(\eta)$. That is, $g$ is a Hessian metric~\cite{Shima-2007}, $(g,\nabla)$ a Hessian structure and $(g,\nabla^*)$ a dual Hessian structure. In differential geometry, $(M,g,\nabla)$ is called a Hessian manifold which admits a dual Hessian structure $(g,\nabla^*)$.
In particular, a Hessian manifold is of Koszul type~\cite{Shima-2007} when there exists a closed $1$-form $\alpha$ such that $g=\nabla\alpha$.

\begin{Remark}
Notice that the potential functions $F$ and $F^*$ are not defined uniquely although the potential functions $\phi$ and $\phi^*$ on the manifold are.
Indeed, consider the generator
$\barF(\theta) =  F(A\theta+b)+ \inner{c}{\theta}+d$
for invertible matrix $A\in\GL(d,\bbR)$, vectors  $b, c\in\bbR^d$ and scalars $d\in\bbR$.
The gradient of the generator $\barF$ is
$\eta=\nabla\barF(\theta) = A^\top  {\nabla F}(A\theta+b)+c$.
Solving the equation $\nabla\barF(\theta)=\eta$ yields the reciprocal gradient 
$\theta(\eta)=\nabla\barG(\eta) = A^{-1}\nabla G\left(A^{-\top} (\eta-c) \right)-b$ from which the Legendre convex conjugate is obtained as  
$\barG(\eta) = \inner{\eta}{\nabla\barG(\eta)}-F(\nabla\barG(\eta))$.
We have $B_F(\theta_1:\theta_1) =   B_{\barF}(\bartheta_1:\bar\theta_2)$
where $\bartheta=A^{-1}(\theta-b)$.
\end{Remark}

It has been shown that a divergence $\calD$ also allows one to define a symplectic structure $\omega$ on a statistical manifold~\cite{Barndorff-1997,zhang2014divergence}.
That is, $\omega$ is a closed non-degenerate $2$-form which is called a symplectic form when exact~\cite{DaSilva-2001}.
The symplectic vector space $(\bbR^{2n},\omega_0)$ viewed as a symplectic manifold has symplectic form $\omega_0=\sum_{i=1}^n \dx^i \wedge \dy_i=-d(\sum_{i=1}^n y_i\dx^i)$ (exact form). 
There is no local invariants but only global invariants on symplectic manifolds (symplectic topology).
That is,  a symplectic structure is flat.

\begin{figure}
\centering
\includegraphics[width=0.65\textwidth]{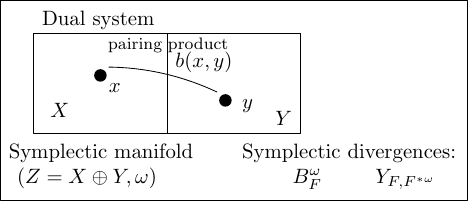}

\caption{Bregman divergences generalized to dual systems $(X,Y,\dualprod{\cdot}{\cdot})$: A symplectic form $\omega$ on the space $Z=X\oplus Z$ is induced by the pairing product. The Bregman divergence on the dual system is then defined as the symplectic Bregman divergence on the symplectic vector space $(Z,\omega)$.}
\label{fig:pairingBregman}
\end{figure}

In this expository paper, we have defined symplectic Fenchel-Young divergences and equivalent symplectic Bregman divergences by following the study of geometric mechanics  reported in~\cite{BuligaSaxce-2017}.
The symplectic Bregman divergence between two points $z_1$ and $z_2$ on a symplectic vector space $(Z,\omega)$ induced by a convex potential function $F:Z\rightarrow\bbR$ is defined by
$$
B_F^\omega(z_1:z_2) = F(z_1)-F(z_2)+\omega(z_1-z_2,\nabla^\omega F(z_2)),
$$
where $\nabla^\omega F$ has been called the symplectic gradient in this paper, and assumed to be the unique symplectic subdifferential at any $z\in Z$,
 i.e., 
$\partial^w F(z)=\{\nabla^\omega F(z)\}$.
Symplectic Bregman divergences are used to define Bregman divergences on dual systems (Figure~\ref{fig:pairingBregman}).
In the particular case of dual system $(X,X,\inner{\cdot}{\cdot})$, we recover ordinary Bregman divergences with composite inner products.

In finite $2n$-dimensional symplectic vector spaces, linear symplectic forms $\omega$ can be represented by symplectic matrices of the matrix group $\Sp(2n)$.
Buliga and de Saxc\'e~\cite{BuligaSaxce-2017} considered geometric mechanics with dissipative terms,
 and stated the following  ``symplectic Brezis-Ekeland-Nayroles principle'' (SBEN principle for short):

\begin{Definition}[SBEN principle~\cite{BuligaSaxce-2017}]
The natural evolution path $z(t)=z_{\rev}(t)+z_{\irr}(t)\in Z$ for $t\in [0,T]$ in a geometric mechanic system with convex dissipation potential $\phi(z)$ minimizes among all admissible paths
$\int_0^T Y_F^\omega(\dot{z}(t),\dot{z}_\irr(t)) \mathrm{d}t$ and
 satisfies
 $Y_F^\omega(\dot{z}(t),\dot{z}_\irr(t))=0$ for all $t\in [0,T]$, where $Y_F^\omega$ denotes the symplectic Fenchel-Young divergence induced by $\phi$, and $z_\rev(t)$ and $z_\irr(t)$ are the reversible and irreversible parts of the particle $z(t)$, respectively.
\end{Definition}

The decomposition of $z=z_\rev+z_\irr$ into two parts can be interpreted as the Moreau's proximation~\cite{Moreau-1965,Rockafellar-1968} associated to the potential function $\phi$:
Indeed, let $F(z)$ be a convex function of $Z=\bbR^{d}$. Then for all $z\in\bbR^{d}$, we can uniquely decompose $z$ as $z=z+z^*$ such that $F(z)+F^*(z^*)=\inner{z}{z*}$ (Fenchel-Young equality) where $z^*=\nabla F(z)$ (see Proposition in \S4 of~\cite{Moreau-1965}). The part $z$ is called the proximation with respect to $F$, and the part $z^*$ is the proximation with respect to the convex conjugate $F^*$.

We may consider the non-separable potential functions $F(z)=F(x,y)=x\, f(y/x)$ which are obtained from the 
perspective transform~\cite{perspectivefunction-2008,Combettes-2018} of arbitrary convex functions $f(u)$ to define symplectic Bregman divergences. 
The perspective functions $F(x,y)$ are jointly convex if and only if their corresponding generators $f$ are convex.
Such perspective transforms play a fundamental role in information theory~\cite{Csiszar-2004} and information geometry~\cite{IG-2016}.

In machine learning, symplectic geometry has been used for designing accelerated optimization methods~\cite{Jordan-2018,shi2019acceleration} (Bregman-Lagrangian framework) and physics-informed neural networks~\cite{matsubara2021symplectic,chen2021neural} (PINNs).

This paper aims to spur interest in either designing or defining symplectic divergences from first principles, and to demonstrate their roles when studying  thermodynamics~\cite{barbaresco2022symplectic} or the learning dynamics of ML and AI systems.

\bibliographystyle{plain}
\bibliography{SympBDBIB}

\end{document}